\documentstyle[prd,aps,epsfig]{revtex}

\def\simgt{\stackrel{>}{{}_\sim}}
\def\gmu{$g_{\mu}-2\ $}
\newcommand\be{\begin{equation}}
\newcommand\ee{\end{equation}}
\newcommand\bea{\begin{eqnarray}}
\newcommand\eea{\end{eqnarray}}
\newcommand\ba{\begin{array}}
\newcommand\ea{\end{array}}
\begin{document}
\draft
\input epsf
\def\la{\mathrel{\mathpalette\fun <}}
\def\ga{\mathrel{\mathpalette\fun >}}
\def\fun#1#2{\lower3.6pt\vbox{\baselineskip0pt\lineskip.9pt
        \ialign{$\mathsurround=0pt#1\hfill##\hfil$\crcr#2\crcr\sim\crcr}}}

\twocolumn[\hsize\textwidth\columnwidth\hsize\csname
@twocolumnfalse\endcsname  

\title{IMPLICATIONS OF MUON $g-2$ FOR SUPERSYMMETRY AND FOR DISCOVERING
SUPERPARTNERS DIRECTLY}
\author{Lisa Everett, Gordon L. Kane, Stefano Rigolin, and Lian-Tao Wang}
\address{\phantom{ll}}
\address{{\it Michigan Center for Theoretical Physics, University of
Michigan, Ann Arbor, MI, 48109}}
%
%
\date{\today}
\maketitle
\begin{abstract}

We study the implications of interpreting the recent muon $g_{\mu}-2$
deviation from the Standard Model prediction as evidence for virtual
superpartners, with very general calculations that include effects of
phases and are consistent with all relevant constraints. Assuming the 
central value is confirmed with smaller errors, there are upper limits 
on masses: at least one superpartner mass is below about 350 GeV (for 
the theoretically preferred value of $\tan\beta=35$) and 
may be produced at the Fermilab Tevatron in the upcoming run, and 
there must be chargino, neutralino, and slepton masses below about 600 GeV. 
In addition, $\tan\beta$ must be larger than about 8. 
\end{abstract}
\pacs{PACS: 12.60.Jv, 13.40.Em   \hskip 0.5 cm
MCTP-01-02 ~~ hep-ph/0102145}
\vskip2pc]

\noindent{\bf Introduction.}

Quantum corrections to the magnetic moments of the electron and the muon
have played major roles historically for the development of basic physics.
The recent report \cite{Brown:2001mg} of a 2.6 standard deviation value of
the muon anomalous moment $a_{\mu}\equiv(g_{\mu}-2)/2$ from its Standard
Model (SM) value $a^{\rm exp}_{\mu}-a^{\rm SM}_{\mu}=426(165)\times
10^{-11}$, assuming it is confirmed as the experiment and SM theory are
further developed, is the first evidence that the Standard Model must be
extended by new physics that must exist on the electroweak scale.  Other
data that imply physics beyond the SM (the matter asymmetry of the
universe, cold dark matter, and neutrino masses) could all be due to
cosmological or very short distance phenomena (though all have
supersymmetric explanations), but a deviation from the SM value of
$g_{\mu}-2$ must be due to virtual particles or structure that exists on
the scale of 100 GeV. Their contribution is as large as or larger than the
effects of the known gauge bosons W and Z, so it must be due to particles
of comparable mass and interaction strength \cite{marciano}.

Taking into account the stringent constraints on new physics from both
direct searches and precision electroweak tests, there are several logical
possibilities to consider.  Presumably the possibility of muon
substructure can be immediately disfavored as effects would already have
been observed in processes involving more highly energetic muons at LEP,
HERA, and the Tevatron. Effects on $g_{\mu}-2$ have been studied in
extensions of the Standard Model such as low energy supersymmetry (SUSY)
\cite{oldgm2,lopez,nath1,moroi,cgw,gm2phase,brignole} or $({\rm 
TeV})^{-1}$ scale extra dimensions \cite{nathextrad,graesser}.  However, 
it has been argued \cite{nathextrad} that the effects due to large extra 
dimensions are
generally small compared to possible effects within SUSY models for
typical parameter ranges.  With this information in hand as well as the
strong theoretical motivation for SUSY due to its resolution of
the gauge hierarchy problem, gauge coupling unification, and successful
mechanism for radiative
electroweak symmetry breaking, we presume that the $g_{\mu}-2$ deviation
is due to supersymmetry and proceed to study it in that context.

The supersymmetry contribution to $g_{\mu}-2$ is not automatically
large. It depends on the superpartner masses and other quantities that are
not yet compellingly predicted by any theory, just as the muon mass itself
is not yet understood. The most important quantity involved besides masses
is called $\tan\beta.$ It is the ratio of the two vacuum expectation
values $v_{1,2}$ of the Higgs fields that break the electroweak symmetry
and give masses to the SM particles; the superpartners also get mass from
these sources as well as from supersymmetry breaking. At the
unification/string scale where a basic effective Lagrangian for a four
dimensional theory is written all particles are massless.  At the
electroweak phase transition the Higgs fields get vacuum expectation
values (VEVs), and the quarks and leptons get masses $m_i=Y_iv_{1,2}$ via
their Yukawa couplings $Y_i$ to the Higgs fields. If the heaviest
particles of each type, the top quark, the bottom quark, and the tau
charged lepton, have Yukawa couplings of order gauge couplings that are
about the same size, as can happen naturally in certain string approaches
and in grand unified theories larger than SU(5) \cite{yukawaunif}, then
the masses and the VEVs are proportional such that the ratio of the VEV
$v_2$ that gives mass to the top quark to the VEV $v_1$ that gives mass to
the bottom quark is $\tan\beta\equiv v_2/v_1 \simeq
m_{{\rm top}}/m_{{\rm bottom}}\simeq35.$

Supersymmetric theories are (perturbatively) consistent for any value of
$\tan\beta$ between about 1 and 50; values of $\tan\beta$ very
near 1 are
already ruled out from direct Higgs searches at LEP \cite{lephiggs}.  A
value of $\tan\beta$ $\simeq35$ has theoretical motivation both from the
unification of the Yukawa couplings just given, and also that 115 GeV is a
natural value for the mass of a Higgs boson if $\tan\beta$ is in this
range \cite{kkw} (this of course is the recently reported value for which
there is evidence from LEP \cite{lep115}). At larger $\tan\beta$ the
supersymmetric contribution to $a_{\mu}$ is essentially proportional to
$\tan\beta,$ as explained below. In minimal supersymmetric theories it is
very difficult to get a Higgs boson mass as large as 115 GeV, so we think
the correlation between the Higgs mass and $g_{\mu}-2$ is significant. 
Since supersymmetry is a decoupling theory, i.e. its contributions
decrease as the superpartner masses increase (see e.g. \cite{herrero}), a
nonzero contribution to \gmu puts an $\mathit{upper}$
limit on the superpartner masses that give the main contributions, the
sleptons (the smuon and muon sneutrino) and the lighter chargino and/or
neutralino.  

In the following we study the one-loop supersymmetric contributions to
$g_{\mu}-2$ with general amplitudes, allowing in particular the full phase
structure of the theory, and we check that the results are consistent with
all relevant constraints. Although \gmu in the context of supersymmetry 
has been studied extensively in the previous literature, if we assume the 
new experimental results will be confirmed and have errors 2-3 times 
smaller soon, an analysis of the data  
yields the first {\it independent upper limits} on slepton and
chargino/neutralino masses, along with a {\it lower bound} on 
$\tan\beta$. We also demonstrate explicitly the effects of the 
relevant phase combination on \gmu in the large $\tan\beta$ regime 
(shown later in Eq.~(\ref{massinsertion1})). In much of the parameter 
space, \gmu only constrains the combination $\tan\beta
\cos\tilde{\varphi}$ (giving the previously unknown result that a nonzero
value for this phase can only
decrease the SUSY effect for a given $\tan\beta$). \\

\noindent {\bf Theoretical Framework.}

The one-loop contributions to $a_{\mu}=(g_{\mu}-2)/2$ in supersymmetric
models include chargino--sneutrino ($\tilde{\chi}^{+}-\tilde{\nu}_{\mu}$)
and neutralino--smuon ($\tilde{\chi}^{0}-\tilde{\mu}$) loop diagrams in
which the initial and final muons have opposite chirality. 
Other contributions are suppressed by higher powers of the lepton masses
and are negligible. As
previously stated, the SUSY contributions to $a_{\mu}$ have been studied
extensively by a number of authors
\cite{oldgm2,lopez,nath1,moroi,cgw,gm2phase}, where the expressions for
these amplitudes can be found.

Note that the majority of these studies assume simplified sets of soft
breaking Lagrangian parameters based on the framework of minimal
supergravity. However, in more general SUSY models the soft breaking
parameters and the supersymmetric mass parameter $\mu$ may be complex.
Several of the relevant phases are severely constrained by
the experimental upper limits on the electric dipole moments (EDMs) of the
electron and neutron, although the constraints can be satisfied by
cancellations \cite{nathedm,bgk,pokorski}. The phases, if nonnegligible,
not only affect CP-violating observables but also can have important
consequences for the extraction of the MSSM parameters from experimental
measurements of CP- conserving quantities, since almost none of the
Lagrangian parameters are directly measured \cite{bk}.  The effects on
$g_{\mu}-2$ due to the phases have recently been
studied in \cite{gm2phase}, where the general expressions for the
amplitudes including phases are presented (but analyzed mainly for small
$\tan\beta$).

The general results of these studies have shown that the SUSY
contributions to $a_{\mu}$ can be large for large $\tan\beta$ and have
either sign, depending on the values of the SUSY parameters.  In
particular, it is well known that for large $\tan\beta$ the chargino
diagram dominates over the neutralino diagram
over most of the parameter space \cite{lopez,nath1,moroi},
and is linear in $\tan\beta$.  
This effect can be seen most easily in the mass insertion approximation,
where the main contribution arises from the chargino
diagram in which the required chirality flip takes place via
gaugino-higgsino mixing rather than by an explicit mass insertion on the
external muon \cite{lopez,nath1,moroi}. In this case, the chargino
contribution to $g_{\mu}-2$ can be written as proportional to:  
\begin{equation} 
\label{massinsertion1} 
a^{\rm \tilde{\chi}^+}_{\mu}\simeq
a^{\rm SUSY}_{\mu} \propto (m^2_{\mu}/\tilde{m}^2)\tan\beta
\cos(\varphi_{\mu}+\varphi_2), 
\end{equation} in which $\varphi_{\mu}$ is
the phase of the supersymmetric Higgs mass parameter $\mu$, and
$\varphi_2$ is the phase of the $SU(2)$ gaugino mass parameter $M_2$; the
reparameterization invariant quantity is
$\tilde{\varphi}\equiv \varphi_{\mu}+\varphi_2$ (note
in the case of zero phases the sign of $a^{\rm SUSY}_{\mu}$ is given in
this limit by the relative sign of $\mu$ and $M_2$
\cite{lopez,nath1,moroi}.)  Also note that $a^{\rm SUSY}_{\mu}$ depends on
$m_{\mu}^2$ because this diagram involves one power of the muon Yukawa
coupling due to the coupling of the right-handed external muon with the
higgsino, and the definition of $a_{\mu}$ is $a_{\mu}=-F_2(0)/2m_{\mu}$
(where $F_2(q^2)$ is the form factor).  This expression can be compared
with the expression for the chargino contribution to the electron EDM in
the mass insertion approximation \cite{pokorski}, as the electric dipole
moment is given by the
imaginary part of $M_2\mu$ while the anomalous magnetic dipole moment is
given by the real part. Therefore, the electron EDM can be obtained from
Eq.~(\ref{massinsertion1}) 
 after replacing
$m_{\mu}\rightarrow m_e$ and $\cos(\varphi_{\mu}+\varphi_2)\rightarrow
\sin(\varphi_{\mu}+\varphi_2)$.
Similar expressions hold for the neutralino
sector \cite{pokorski,moroi}; while they are generally suppressed due to
the smaller neutral current coupling, they can be relevant for cases in
which $m_{\tilde{l}}, M_1 \ll M_2,\mu$.

The fact that $a^{\rm SUSY}_{\mu}$ may have either sign at first may seem
counterintuitive, given the well-known result \cite{ferrara,bg,gm2phase}
that $a^{\rm SUSY}_{\mu}$ exactly cancels $a^{\rm SM}_{\mu}$ in the
unbroken SUSY limit, with the cancellations taking place between the
chargino and the W, the massless neutralinos and the photon, and the
massive neutralinos and the Z. (The general statement \cite{ferrara} is
that any magnetic-transition operator vanishes in this limit, due to the
usual cancellation between fermionic and bosonic loops in SUSY theories.)
As $a^{\rm SM}_{\mu}$ is known to be positive \cite{gm2SM}, $a^{\rm
SUSY}_{\mu}$ is negative in this limit.  However, the limit with unbroken
SUSY but broken electroweak symmetry requires the supersymmetric Higgs
mass parameter $\mu=0$ and $\tan\beta=1$, as can be seen from the Higgs
potential when the soft breaking parameters are zero:
$V=|\mu|^2(v_1^2+v_2^2)+ (g^2+g'^2)(v_2^2-v_1^2)^2/2$. At low (but $\simgt
1$) values of $\tan\beta$ and nonzero $\mu$, the chargino and neutralino
contributions are comparable in magnitude but opposite in sign; however 
the neutralino diagram dominates as the parameters deviate from the 
unbroken SUSY limit since the contribution from the (nearly) massless
neutralinos is much larger than that of the massive neutralinos and
charginos (recall this contribution cancels the much larger photon
contribution in the exact SUSY limit).  
At larger values of $\tan\beta$ the chargino diagram
begins to dominate and the sign of $a^{\rm SUSY}_{\mu}$ can flip depending
on the relative sign (or phase) of $\mu$ and $M_2$.  In the traditional
convention in which $M_2$ is chosen to be real and positive, the sign of
\gmu is given by the sign of $\mu$.  We pause here to comment on the
sign of $\mu$ as it relates to the $b\rightarrow s \gamma$ decay
\cite{bg,bsgamma}.
In the 
literature it is often claimed that the constraints on the SUSY parameter
space due to the requirement of cancelling the various SUSY contributions
to the $b\rightarrow s \gamma$ branching ratio place a strong constraint
on the sign of $\mu$.  In SUSY models with general phases, the branching
ratio actually involves a reparameterization invariant combination of
phases $\varphi_{\mu}+\varphi_{A_t}$ (which reduces to the relative sign 
for zero phases).  The
relative sign from $b\rightarrow s \gamma$ thus is not the same
combination as that constrained by \gmu; however, in the usual conventions
with
$M_2$ and $A_t$ real and positive, both processes favor positive $\mu$.\\

\noindent {\bf Results and Discussion.}

We calculated the complete one--loop MSSM contribution to $g_{\mu}-2$,
taking  into account the possibility of nontrivial CP-violating phases for
the $\mu$ term and the soft breaking parameters
(see \cite{stef} for general formulae and conventions).
We have made a few simplifying assumptions
which do not significantly impact 
our general
conclusions. First, we have assumed that the masses of
the charginos and neutralinos are heavier than 100 GeV. This assumption
will be easily verified for the charginos as soon as LEPII will report its
final results on new searches at $\sqrt{s}
= 208$ GeV. LEPII will not be able to provide such a limit on 
the neutralinos, but the neutralino contribution is usually small 
compared to the charginos so the assumption that neutralinos are
not very light does not affect the upper bounds we put on superpartner
masses (it does affect the $\tan\beta$ bound, as explained below).
In addition, we assumed a common slepton mass $m_{\tilde l} > 100$
GeV for the left and right smuons and for the muon sneutrino. 
For very
light masses that isn't a good approximation, but it does not change our 
numerical results.  
We assumed also
that $|\mu| \tan \beta >> |A_{\mu}|$, which is a
reasonable assumption for $\tan \beta >$ 3. Moreover, as the smuon
mass matrix enters only in the suppressed neutralino contribution, the
details of the charged slepton mass matrix are almost irrelevant in the
numerical analysis (except at certain exceptional points in parameter
space). Thus, the most important parameter in the slepton sector is the
sneutrino mass, which is likely to have a LEPII lower limit.
However, as 
the sneutrino mass enters only in the loop integrals as a suppression, 
relaxing this assumption is not going to change our general conclusions.

In Figure 1 we show the $m_{\tilde{\chi}_1}-m_{{\rm slepton}}$ range 
allowed at 1 $\sigma$ by the \gmu measurement for four different
values of $\tan\beta$ ($\tan \beta$= 10, 20, 35 and 50), where
$m_{\tilde{\chi}_1}$ denotes the lightest chargino or neutralino, and
$m_{{\rm slepton}}$ denotes the smuon or muon sneutrino. The \gmu 
data will provide the first phenomenological {\em upper} limit on 
superpatner masses. In order to learn what limits could arise when the 
data already in hand is analyzed (including data on the hadronic 
contribution), we present results allowing 1 $\sigma$ deviations from 
the present central values. Thus these are the strongest results that 
could be implied by the data (unless the central value increases even 
more). The region above the lines is excluded because the masses are 
too heavy to account for the experimentally observed $\delta a_\mu$ 
difference. These regions are obtained for zero phases.
In the regions of parameter space in which the
chargino diagram dominates, \gmu depends on $\tan\beta \cos(\varphi_\mu +
\varphi_2)$ (see Eq.~(\ref{massinsertion1})), such that nonzero phases
only
decrease \gmu for a given $\tan\beta$.  In addition, our combined analysis
for the electron EDM shows that in this region $\varphi_{\mu}+\varphi_2$
is severely constrained to be less than $10^{-2}$,
due to the fact that cancellations are more difficult to achieve for
large $\tan\beta$ (and two-loop contributions which we have neglected
here may become important \cite{ckp}). In certain exceptional points of
parameter space in which the neutralino and chargino diagrams are
comparable (i.e. with light sleptons and $M_1 \ll m_2,\mu$), the EDM
cancellations must be reconsidered taking large $\tan\beta$ into
account, which we defer to a future study.  

From Figure 1, important constraints on the MSSM
parameter space can be obtained. There is a maximum range allowed for the
lightest chargino, neutralino, and slepton masses. For perhaps the most
interesting case of $\tan \beta = 35$, values of $m_{\tilde{\chi}_1} >$ 550 
GeV and ${m}_{\tilde{l}} >$ 650 GeV are disfavored by \gmu measurements 
(note if one is near the limit the other is much
lighter). For lower $\tan \beta$ allowed masses are always
smaller. ``Effective supersymmetry'' scenarios \cite{effsusy}, with
TeV mass first and second generation squarks and sleptons, are ruled out.

Further, the \gmu measurement implies 
a lower bound for $\tan\beta \simgt $ 8 (note nonzero phases do not
affect this lower bound). Lower values of $\tan \beta$ 
give too small a contribution to \gmu; however, there 
is a small corner of parameter space for ultralight neutralinos (with 
masses of $\approx O(50 \,{\rm GeV})$) where 
$\tan\beta$ can be as little as about half of our limit. 
As improved measurements become available, \gmu will determine $\tan\beta$
with increased precision. Measuring $\tan\beta$ is extremely
difficult at hadron colliders \cite{bk}, yet $\tan\beta$ is an extremely
important parameter for supersymmetry predictions and tests. 
To obtain a large
$g_{\mu}-2$ large $\tan\beta$ is necessary, and since the size is 
essentially  proportional to $\tan\beta$ it is immediately approximately 
known. It can then be determined
accurately when a few superpartner masses and the soft
phases (which are constrained from EDMs) 
are known. \\

\noindent {\bf Summary.}

Because the reported \gmu number is larger than the Standard Model
prediction by an amount larger than the W and Z contributions, it implies
several significant results. We presume the effect arises from
superpartner loops; the chargino--(muon)sneutrino loop typically
dominates, with the neutralino--smuon relevant in certain restricted
regions of parameter space. Then, in approximately decreasing order of
interest, 
\begin{itemize}

\item One superpartner, either a chargino, neutralino, or a slepton, has
to be lighter than about 350 GeV (for the theoretically motivated value 
of $\tan\beta=35$; see Figure 1 for a more precise number).

\item The heavier one of the lightest chargino or sneutrino has to be
lighter than about 600 GeV, so models with heavier sleptons are
disfavored.

\item $\tan\beta$ has to be larger than about 8 (in corners of 
parameter space with ultralight neutralinos, $\tan\beta$ can be lower).  
This large 
$\tan\beta$
is sufficient to obtain a Higgs boson mass of about 115 GeV, and also
implies a number of interesting phenomenological consequences (e.g.
\cite{largetanbeta}).


\end{itemize}

The \gmu measurement is the first data to establish a firm {\it upper}
limit on any superpartner masses.  Over most of the allowed masses, the
superpartners will be produced at Fermilab in the upcoming run.

\begin{figure}[hbt]
\label{fig1}
\vskip -0.5cm
\hskip -1cm
\centerline{
\epsfig{figure=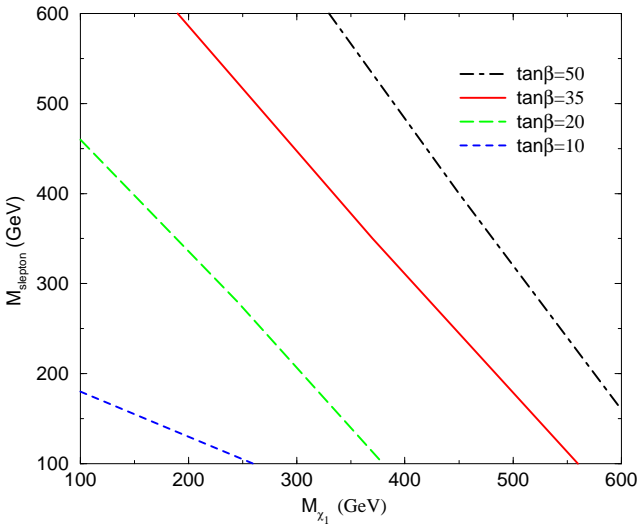,height=8truecm}}
\caption{\footnotesize In this figure $m_{\chi_1}$ denotes the lightest
chargino or neutralino, and $m_{{\rm slepton}}$ the lightest smuon or
muon sneutrino. The regions above a given $\tan \beta$ line are
excluded. We require agreement with experiment within $1 \sigma$ (see
text). Below $\tan \beta \approx 8$ no allowed region remains. $\tan
\beta$, the ratio of the two Higgs vacuum expectation values, is defined
in the text; approximate Yukawa coupling unification suggests $\tan
\beta \sim 35$. Thus the figure implies related upper limits on the
lightest chargino/neutralino and slepton masses.}
\end{figure}

\end{document}